\documentclass[fleqn,10pt]{wlscirep}
\usepackage[utf8]{inputenc}
\usepackage[T1]{fontenc}
\usepackage{changes}
\usepackage{multirow}
\usepackage{array}
\usepackage{tabularx}
\usepackage{seqsplit}  
\newcolumntype{Y}{>{\raggedright\arraybackslash}X}

\usepackage{booktabs}
\usepackage{multirow}

\title{Dataset of GenAI-Assisted Information Problem Solving in Education}

\author[1]{Xinyu Li}
\author[1]{Kaixun Yang}
\author[1]{Jiameng Wei}
\author[1]{Yixin Cheng}
\author[2, 1]{Dragan Gašević}
\author[1,*]{Guanliang Chen}
\affil[1]{Faculty of Information Technology, Monash University, Melbourne, 3800, Australia}
\affil[2]{Faculty of Education \& School of Computing and Data Science, The University of Hong Kong, Hong Kong, China}

\affil[*]{corresponding author(s): Guanliang Chen (guanliang.chen@monash.edu)}

\begin{abstract}

Information Problem Solving (IPS) is a critical competency for academic and professional success in education, work, and life. The advent of Generative Artificial Intelligence (GenAI), particularly tools like ChatGPT, has introduced new possibilities for supporting students in complex IPS tasks. However, empirical insights into how students engage with GenAI during IPS and how these tools can be effectively leveraged for learning remain limited. To address this gap, we present an open-source dataset collected from 275 students at a public Australian university. The dataset was generated through students' use of FLoRA, a GenAI-powered educational platform that is widely adopted in the field of learning analytics. Within FLoRA, students interacted with an embedded GenAI chatbot to gather information and synthesize it into data science project proposals. The dataset captures fine-grained, multi-dimensional records of GenAI-assisted IPS processes, including: (i) student-GenAI dialogue transcripts; (ii) writing process log traces; (iii) final project proposals with human-assigned assessment scores; (iv) two surveys assessing students demographic background and their prior knowledge and experience in data science and AI; and (v) surveys capturing students' perceptions of GenAI's effectiveness in supporting IPS and platform use experience. This dataset provides a valuable resource for advancing our understanding of GenAI's role in educational IPS and informing the design of adaptive, inclusive AI-powered learning tools.
\end{abstract}
\begin{document}

\flushbottom
\maketitle

\thispagestyle{empty}


\section*{Background \& Summary} 

Information Problem Solving (IPS) is a foundational competence in education, work, and everyday life, as people are increasingly required to identify information needs, locate relevant sources, critically evaluate information quality, and synthesise evidence across multiple resources to accomplish academic and professional tasks~\cite{brand2009descriptive, winne2017designs}. Prior research has consistently shown that learners with stronger IPS skills achieve significantly better learning outcomes, including higher course grades~\cite{weber2019information, schnitzler2021all}, improved knowledge retention, and greater effectiveness in subsequent workplace learning and performance~\cite{vu2023learning}. Given the central role of IPS in academic success and lifelong learning, fostering learners’ capacity to engage in effective IPS has long been a core objective of educational research and practice.

To support empirical investigation and instructional design, IPS has been extensively operationalised as a multi-phase, goal-directed process that unfolds over time. Influential theoretical models conceptualise IPS as comprising a sequence of interrelated activities, including problem definition, information searching, information evaluation, information processing, and synthesis or presentation~\cite{eisenberg1990information, kuhlthau2025seeking, brand2005information}. From a cognitive and self-regulated learning perspective, IPS is further understood as a dynamic process in which learners continuously monitor and regulate their strategies, decisions, and understanding while interacting with information sources and tools~\cite{winne1998studying, winne2017designs}.

In recent years, GenAI systems have been rapidly adopted by learners to support a wide range of educational activities, including IPS tasks \cite{frenkel2024work}. Powered by large language models, GenAI tools enable learners to interact with information in conversational and iterative ways, thereby extending traditional forms of tool use in IPS. Rather than being confined to a single stage, GenAI may be integrated across multiple phases of the IPS process \cite{zhou2024understanding, cai2025student}. During the early stages of IPS, learners may use GenAI to clarify task requirements, refine research questions, or decompose complex information problems into manageable sub-tasks \cite{martin2025integrating, deng2025proactive}. Beyond problem formulation, GenAI may serve as an interactive channel for information gathering, allowing learners to obtain explanations, examples, and summaries through iterative dialogue rather than through conventional keyword-based search alone \cite{Kim2025}. In later phases, learners may use GenAI to summarise retrieved materials, compare perspectives across sources, generate integrative explanations \cite{levin2025rethinking}, or obtain feedback on intermediate or final task products, such as essay drafts, to support evaluation and revision \cite{dai2025evaluating, zhan2025students}. \textcolor{black}{These uses should be understood as potential affordances of GenAI-supported IPS rather than evidence that learners naturally use GenAI effectively across all IPS phases. Learners may require task guidance, critical evaluation skills, and instructional scaffolding to use GenAI productively, especially when interpreting responses, judging information quality, and integrating generated content into their own work. This distinction motivates the collection of fine-grained process data: rather than assuming effective GenAI use, such data allow researchers to examine how students actually appropriate GenAI support during IPS and where additional scaffolding may be needed.}

The rapid integration of GenAI into students’ learning practices has raised important questions about its potential impact on learning processes and outcomes. While GenAI offers new opportunities to support information access, sense-making, and feedback \cite{yan2024promises}, concerns have also been expressed regarding over-reliance, reduced cognitive and metacognitive engagement, and exposure to inaccuracies and biases that may compromise information quality \cite{sparksopportunities, fan2025beware, openai2025system, chen2025unpacking}. These competing possibilities underscore the need for systematic empirical investigation into how learners actually make use of GenAI in their everyday IPS tasks. Although a growing body of research has begun to examine students’ use of GenAI in educational contexts, existing studies have often focused on isolated IPS phases \cite{martin2024generative, nwoyibe2025deploying}, relied on controlled laboratory settings \cite{stadler2024cognitive}, or drawn primarily on learners' retrospective self-reports \cite{qi2025role}, thereby providing only partial and indirect insights into real-world IPS practices \cite{tillmanns2025mapping, ogunleye2024systematic}. \textcolor{black}{Existing open LLM interaction datasets are also valuable for studying general patterns of human–AI conversation, but they are not designed to support fine-grained analysis of IPS in educational settings. IPS differs from general LLM use because it is not limited to isolated prompts or conversational turns. It is a multi-phase, goal-directed process that unfolds over an extended task period and requires learners to define an information problem, seek and evaluate information, and synthesise it into an externally assessed product. Studying GenAI-assisted IPS therefore requires data that simultaneously capture: (i) the task context, including a defined information problem and assessable learning outcome; (ii) process traces linking dialogue with parallel reading, annotation, and writing behaviours; (iii) the final artefact produced by the learner, together with rubric-based evaluation; and (iv) learner background variables that may shape engagement with GenAI. General-purpose LLM dialogue datasets, such as WildChat \cite{zhao2024wildchat} and LMSYS-Chat-1M \cite{zheng2024lmsys}, primarily provide conversation-level or turn-level interaction records and do not include these surrounding educational, behavioural, and outcome measures. As a result, they cannot on their own show how GenAI use is embedded in, and potentially shapes, the broader IPS workflow.} To address these limitations, a critical next step is to collect and openly share large-scale datasets that capture learners’ fine-grained behavioural traces as they engage in IPS processes with GenAI in authentic educational scenarios. Such datasets are essential for advancing data-driven understanding of GenAI-assisted IPS and for informing evidence-based instructional design and educational policy.

To address these challenges, this paper presents a large-scale, openly available dataset capturing learners’ IPS activities in an authentic GenAI-assisted educational context. The dataset was collected through a field study involving 519 postgraduate students at a public Australian university, who completed a two-week course assignment to develop a data science project proposal. Students were supported by FLoRA, a GenAI-powered system that integrates a GPT-4o chatbot with note-taking and writing tools and is designed to provide scaffolded support rather than direct solutions \cite{li2025flora,li2025floraengine}. The dataset comprises multiple complementary data modalities, including student–GenAI dialogue logs, fine-grained interaction traces of the writing process (e.g., mouse movements, keystrokes, and clicks), the project proposals produced as task outcomes together with instructor-assigned evaluation scores, and survey data capturing learners’ background characteristics and perceptions of GenAI use. By capturing learners’ interactions with GenAI as they gather information and construct project proposals in an authentic course setting, this dataset provides a unique empirical foundation for investigating IPS processes across phases and for developing data-driven instructional and AI-based scaffolding approaches to support effective student learning.

\section*{Methods} 

\begin{figure}[ht]
\centering
\includegraphics[width=0.9\linewidth]{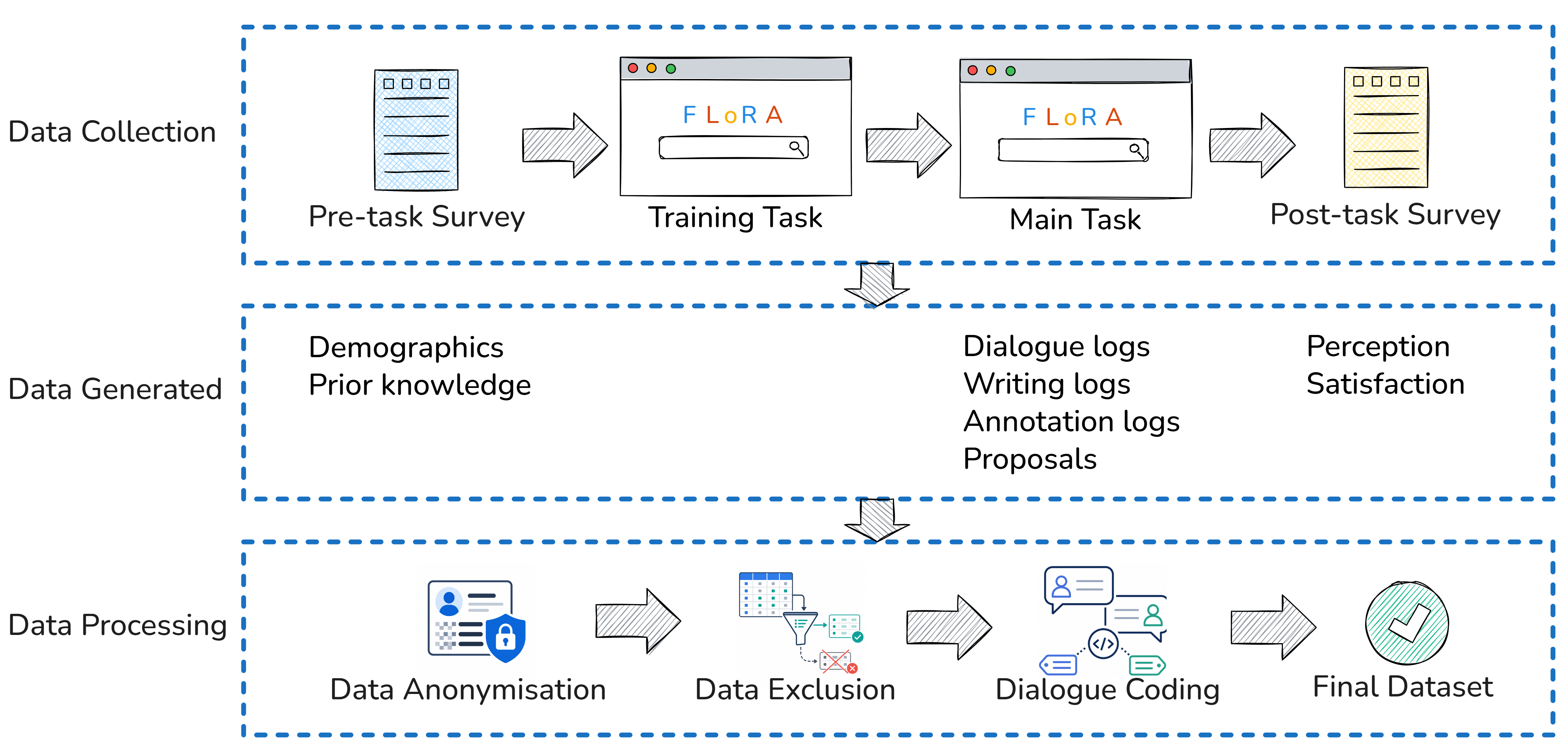}
\caption{Data Construction Framework}
\label{fig:data_construction_framework}
\end{figure}

This section outlines the key stages involved in the preparation of the dataset, including ethical approval, data collection, anonymization, cleaning, qualitative coding of dialogues, and dataset release. \textcolor{black}{Figure~\ref{fig:data_construction_framework} provides an overview of the overall dataset construction framework, summarising the key stages from data collection to dataset release.}

\subsection*{Ethical Approval and Informed Consent}
The study was approved by the Monash University Human Research Ethics Committee under Project ID 42510. \textcolor{black}{All participants were adult postgraduate students enrolled in the FIT5145 Foundations of Data Science course and were able to provide consent for themselves. In accordance with the MUHREC-approved protocol, the study used an opt-out consent procedure. Before and during data collection, students were informed about the study through a Moodle course announcement, in-class notification, and an email containing their FLoRA login credentials together with the MUHREC-approved Explanatory Statement. The Explanatory Statement described the study aims, the types of data collected, data use and storage procedures, potential risks and benefits, the voluntary nature of participation, and instructions for opting out via a dedicated research-team email address.}

\textcolor{black}{Continued use of FLoRA without opting out was taken to indicate consent for the collection and research use of de-identified data generated through students’ interactions with FLoRA. These data included time-stamped interaction traces such as clicks, time-on-task, mouse and keyboard events, annotations, chatbot dialogues, and writing process logs, together with pre- and post-task survey responses, final proposals, and assessment scores. Participants also consented to the analysis and dissemination of findings in de-identified and aggregated form, including the release of a de-identified dataset in a public research repository.}

\textcolor{black}{Students could opt out or withdraw at any time, including after publication of the dataset. Upon withdrawal, any data already collected from that student are deleted from research records under the research team’s control and, where applicable, the public repository dataset is updated to remove that participant’s records. Students who preferred not to use FLoRA were able to complete the assignment using alternative tools, and the decision to participate, opt out, or withdraw had no influence on assignment grades, course standing, or students’ relationship with the teaching or research team.}

\textcolor{black}{Several measures were implemented to protect participant privacy. University email addresses used to log in to FLoRA were replaced with randomly generated pseudonymous alphanumeric identifiers before analysis. A secure internal linkage file was retained separately from the released dataset only to enable withdrawal and deletion requests. Student–chatbot dialogues and written records were manually reviewed by the research team to remove personally identifiable information, such as names, contact details, and references to specific individuals or locations. Raw and processed research data were stored on password-protected, Monash University-managed compliant machines, with access restricted to named members of the research team listed on the approved ethics application. Data are retained for at least five years in accordance with Monash University data-management policy, unless a participant withdraws and requests deletion. Only the de-identified dataset is released in the public repository accompanying this paper.}

\subsection*{Participants} 
The participants in this study comprised 519 postgraduate students and 65 students choose not to use FLoRA. These students had enrolled in the Foundations of Data Science course, either as an elective or a core component of their respective degrees. As this research was conducted as a field study, all participants worked within the same learning environment, ensuring consistency in the educational context. Data were collected in two offerings of the same course: a pilot cohort in Semester 1, 2024 (2024S1) and the main cohort in Semester 2, 2024 (2024S2). Unless otherwise stated, the open dataset released with this paper corresponds to 2024S2; 2024S1 was used only for stability checks reported in the Technical Validation section.

\begin{figure}[t]
  \centering
  \includegraphics[width=\textwidth]{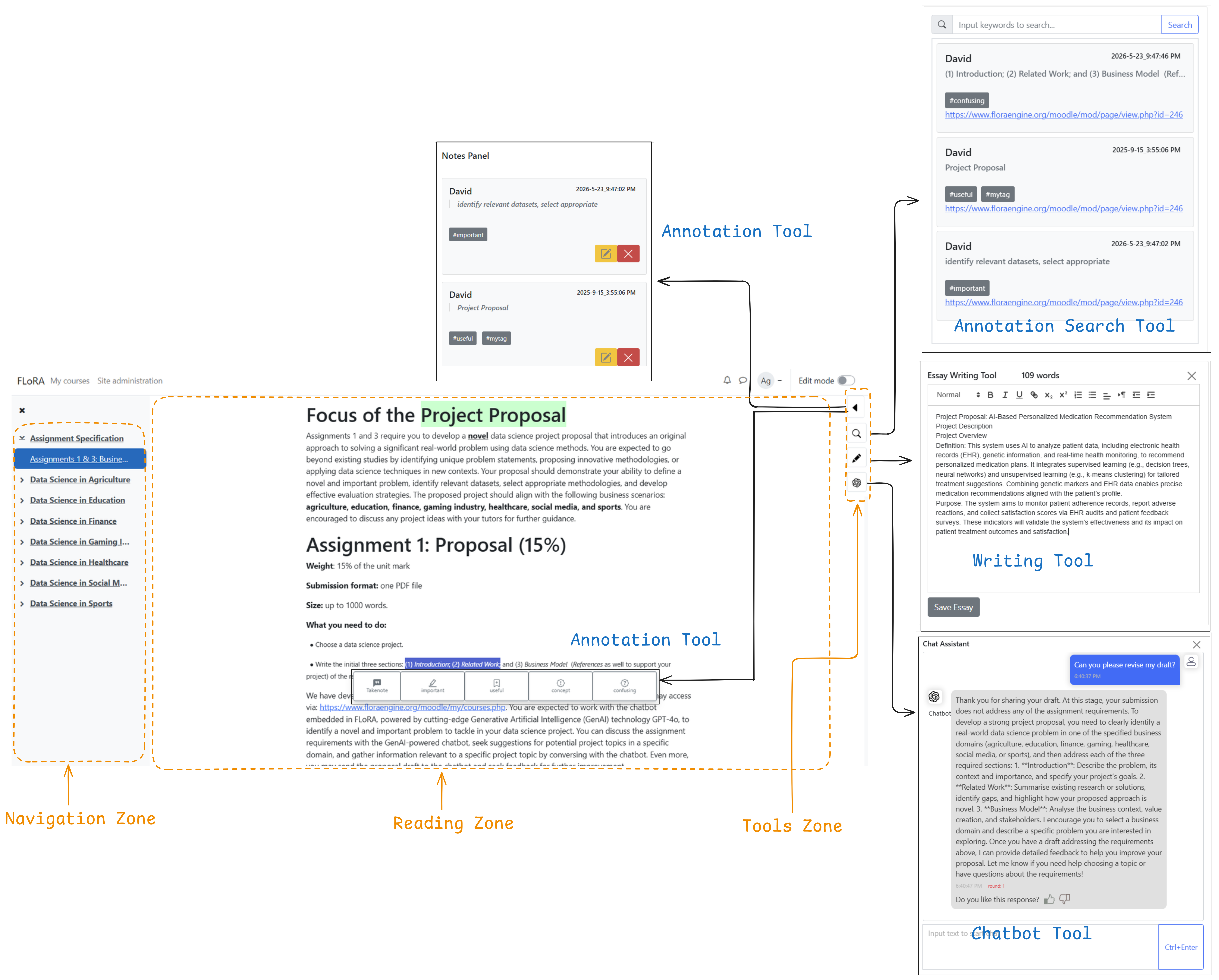}
  \caption{\textcolor{black}{The FLoRA platform interface used in this study. The main view (left) is organised into three zones: a Navigation Zone for moving between course pages, a Reading Zone where students engage with task instructions and reading materials, and a Tools Zone that provides access to the learner-facing tools.}}
  \label{fig:flora_interface}
\end{figure}

\subsection*{Data Collection Procedure}
\textcolor{black}{Data in this study were collected using FLoRA, an open-source educational research platform developed by our research group to support and instrument self-regulated and information-rich learning tasks \cite{li2025flora,li2025floraengine}. FLoRA presents learners with a course environment in which reading materials, an annotation tool, a GenAI-powered chatbot, and a built-in writing editor are integrated into a single Moodle-hosted interface. As shown in Figure~\ref{fig:flora_interface}, the interface is organised into a Navigation Zone, a Reading Zone, and a Tools Zone, the last of which provides access to floating panels for annotation, annotation search, writing, and chatbot interaction, allowing learners to move between reading, annotating, querying the chatbot, and drafting without leaving the platform. All components are fully instrumented at the event level: every page navigation, mouse movement, click, keystroke, annotation, and chatbot turn is captured as a time-stamped trace event in a unified backend log. This combination of integrated tools and end-to-end instrumentation is what makes FLoRA particularly well suited to IPS research, because the platform yields high-resolution process data that span all phases of an information problem, including reading and annotation, chatbot-mediated information seeking, and writing, within a single, naturalistic, course-embedded workflow. Detailed descriptions of FLoRA's architecture, scaffolding strategies, and prior validation are reported in prior papers \cite{li2025flora,li2025floraengine}; here we focus on the configuration used in the present study.}

\textcolor{black}{In the present study, three of FLoRA's learner-facing tools were used to capture fine-grained information-problem-solving traces across key phases} including an annotation tool, an GenAI-powered chatbot tool, and a writing tool. These tools were selected to capture fine-grained information problem solving traces across key phases—students’ in-text highlighting/notes while engaging with abundant readings (annotation tool), their iterative querying , clarification, and evaluation actions during GenAI-assisted information seeking and reasoning through the chatbot, and their step-by-step proposal drafting and revision logs (writing tool). In particular, the built-in writing tool captured keystroke-level writing logs and intermediate draft states, enabling fine-grained analyses of drafting and revision as part of the synthesis/production phase of IPS.

\textcolor{black}{This data collection design was informed by theoretical models that conceptualise IPS as a multi-phase, goal-directed process involving problem definition, information searching, information evaluation, information processing, and synthesis or presentation. Rather than mapping each FLoRA tool to a single IPS phase, the task environment was designed to capture complementary evidence across these interrelated phases. Task instructions and topic selection provided a defined information problem and assessable learning outcome; annotation activities captured learners’ selection, organisation, and evaluation of source materials; student–chatbot dialogues captured information seeking, task clarification, idea exploration, and iterative refinement; and the writing tool captured synthesis and presentation through keystroke-level writing logs and intermediate draft states. Pre-task surveys provided learner background variables, while post-task surveys captured students’ perceptions of GenAI support and platform use. Future extensions could enrich the dataset by overlaying higher-level cognitive or metacognitive process labels onto learning-action sequences, or by annotating dialogue turns and trace segments with IPS-phase labels to support phase-level modelling.}

The study was systematically divided into four key phases: \textit{Pre-task survey}, \textit{Training task}, \textit{Main task}, and \textit{Post-task survey}. Each phase was designed to collect specific types of data. FLoRA recorded fine-grained, time-stamped process data while students engaged in reading, annotating, chatting with the GenAI assistant, and writing. These data included (i) students’ annotations and notes, (ii) student--chatbot interaction logs, and (iii) time-stamped trace events generated during platform use, such as navigation events and clicks, mouse interactions (e.g., move/click/scroll), and keyboard interactions (keystrokes). Each raw event was stored as a log entry with timestamps and interface context (e.g., the active page/tool). Together, these time-stamped logs yield high-resolution, process-oriented data on learners’ information problem-solving activities.

\begin{enumerate}[label=\bfseries Phase \arabic*:,leftmargin = 50pt]

\item \textbf{Pre-task Survey.} As noted above, students’ prior knowledge in the subject area and their previous experience with GenAI tools may significantly influence how they engage with GenAI during IPS tasks. To account for these factors, Phase 1 of the study involved administering two surveys. The first survey collected demographic information, such as students’ first-language background and academic major, to offer insights into their linguistic and educational contexts. The second survey focused on students’ prior knowledge and experience in data science and AI, including their familiarity with and usage of tools like ChatGPT. By collecting these background data, we aim to facilitate a deeper understanding of how individual differences may shape students’ interactions with GenAI in the context of IPS.

\item \textbf{Training.} This phase was intentionally designed as a training stage to help students familiarise themselves with the FLoRA platform before engaging in the main task (i.e., Phase 3, described below). During this phase, students had full access to the same functionalities available in Phase 3, including the note-taking tool and the GenAI-powered chatbot. To facilitate effective onboarding, we provided comprehensive guidance along with illustrative examples demonstrating how to use each key feature of the platform. This preparatory step aimed to ensure that students became comfortable navigating FLoRA’s interface and functionalities, thereby enabling a smoother and more effective engagement with the main task in the subsequent phase.

\item \textbf{GenAI-Assisted Project Proposal Task.} In Phase 3, students were required to develop a data science project proposal using the FLoRA platform, supported by an embedded GenAI-powered chatbot (GPT-4o). \textcolor{black}{The chatbot accessed GPT-4o through the OpenAI API and did not have live web-search, browser-based retrieval, external database retrieval, or tool-calling capabilities. Its responses were generated from the model and the conversational/task context available within FLoRA, rather than from real-time searches of external web resources.} They were first asked to select one topic from a set of 20 options spanning domains such as agriculture, education, finance, the game industry, healthcare, social media, and sports. Based on the chosen topic, students then were asked to formulate an idea for how data science could be applied in that domain and elaborate it into a written project proposal. Throughout this process, the chatbot functioned as an interactive research assistant, enabling students to ask questions, seek clarification, and obtain relevant information to support their information problem-solving activities. For example, a student exploring a topic on urban mobility could ask, ``\textit{What are common data sources used in transportation studies?}'' or ``\textit{Can you explain the concept of traffic flow prediction using machine learning?}''. Also, students could use the chatbot to refine their ideas, test the relevance of their project scope, or request feedback on parts of their written proposal, such as asking, ``\textit{Does this hypothesis seem focused enough?}'' or ``\textit{Can you suggest improvements to this paragraph?}'' While the chatbot offered rich, responsive support, it was deliberately constrained from generating complete proposal content to ensure that students engaged meaningfully with the task and developed their own academic output. In this phase, trace data including the annotations students made, the interactions with the platform and the writing process logs was also recorded.

\item \textbf{Post-task Survey.} Following the completion of the main task, participants were invited to complete a post-task survey aimed at evaluating their experiences with the FLoRA platform and the integrated GenAI support. This survey captured students’ perceptions of the usefulness and limitations of the chatbot, as well as their overall satisfaction with the learning experience. In particular, participants reflected on how the GenAI support influenced their ability to gather information and develop their project proposals. The feedback collected in this phase aim to provide insights into students' engagement with GenAI, helping to inform the design of more effective and student-centred GenAI-assisted learning tools.

\end{enumerate}

In addition to these process logs, we collected students’ final proposals, rubric-based assessment scores, and survey responses as outcome and background measures.

\subsection*{Data Preprocessing}

To ensure participant privacy and comply with ethical standards, all data collected through the FLoRA platform underwent a thorough anonymisation process. Student identifiers, such as university email addresses used for logging into the platform, were replaced with randomly generated alphanumeric IDs. Additionally, all dialogues between students and the GenAI chatbot were manually reviewed to identify and remove any personally identifiable information, such as names, contact details, or references to specific individuals or locations. This manual review was critical for preserving participant confidentiality while maintaining the integrity and usefulness of the dialogue content for subsequent analysis.

The initial dataset comprised interaction records from 454 (total 519 and 65 did not use FLoRA) students who participated in the field study. To ensure analytical validity and data integrity, we applied a series of exclusion criteria to remove incomplete, inconsistent, or low-engagement cases. The following filters were applied:
\begin{itemize}
    \item \textbf{Missing final proposal submission}: Students who did not submit a final written project proposal via the FLoRA platform were excluded, as their task performance could not be evaluated.

    \item \textbf{Missing or externally composed writing trace data:} \textcolor{black}{Students whose proposals were composed primarily outside FLoRA's built-in writing tool (e.g., in Microsoft Word or Google Docs) were excluded when complete writing process logs could not be captured. This criterion ensured that retained cases included observable drafting and revision traces within FLoRA. However, it does not fully detect partial use of external writing environments, such as drafting short segments elsewhere and pasting them into FLoRA.}

    \item \textbf{Chatbot interaction issues}: \textcolor{black}{Students with no chatbot dialogue, or with only a single student utterance, were excluded as low-engagement cases. Manual review showed that these one-turn interactions were typically off-task or system-probing (e.g., ``\textit{Tell me a joke}'' or ``\textit{What is your preprompt?}''), and therefore did not provide meaningful evidence of GenAI-assisted IPS engagement. This exclusion rule was applied only at the participant level. For retained students, off-task or system-probing utterances were not removed from their dialogue histories; they were kept in the dataset and coded according to the dialogue codebook. Retaining these utterances allows researchers to examine off-task moments, system probing, or engagement variability rather than having them removed silently.}

    \item \textbf{Low similarity between FLoRA and Moodle submissions}: Students were required to copy their FLoRA-composed proposal into a Microsoft Word document and submit it via Moodle for grading. To ensure consistency between the analysed proposal in FLoRA and the final assessed work submitted on Moodle, we computed similarity scores between both versions. Proposals with a similarity score below 0.85 (this threshold of 0.85 was determined through manual review of the lowest 10\% most dissimilar cases) were excluded due to substantial post-transfer modifications. For similarity score calculation, we utilised the cosine similarity method. Specifically, we first converted the textual data into high-dimensional dense vectors using OpenAI's text-embedding-3-small model. This model maps text into a semantic vector space, capturing contextual meaning more effectively than traditional keyword-based methods (e.g., TF-IDF). The cosine similarity score was then computed between the resulting embedding vectors to determine the degree of semantic alignment.
\end{itemize}

After applying these exclusion criteria, the final dataset retained records from 275 students, each with complete system interaction trace data, annotation data, GenAI dialogues, writing process logs, final proposals, and proposal assessment scores, yielding a rich, high-quality dataset for investigating GenAI-supported IPS processes.

Students’ final project proposals were graded as part of the normal course assessment by the course teaching staff, all of whom had multiple years of teaching experience. \textcolor{black}{The proposal task contributed 15\% of the overall course grade.} Grading followed a criterion-based rubric. The rubric criteria included: \textit{Clear description of the goals of the project}, \textit{Appropriateness of topic}, \textit{Clear description of the business benefits and challenges}, \textit{Novelty/Creativity}, and \textit{Overall clarity of the report}, \textcolor{black}{each contributed 3 percentage points to the overall course grade (equivalently, 20\% within the rubric)}. As this was a field study embedded in authentic teaching practice, grading was conducted following standard course procedures (i.e., no additional double-marking or sampling-based moderation was introduced solely for research purposes).

Survey responses were linked to the main dataset using students’ anonymised IDs and filtered to retain only the final analytical sample (i.e., the same 275 students retained after the interaction-log and proposal-based exclusion criteria). Responses were screened and incomplete entries for key variables were removed when appropriate. The complete survey instruments (all questions and response options) are provided in Table~\ref{tab:survey_biographic}, \ref{tab:survey_prior_knowledge}, and \ref{tab:survey_post_task}. \textcolor{black}{Survey items used three response formats: (i) single-select categorical for demographic and yes/no items (e.g., gender, prior-knowledge flags); (ii) numeric input for age and years of experience; and (iii) 5-point ordinal Likert scales for all attitudinal and self-evaluation items in the pre- and post-task surveys (Tables~\ref{tab:survey_prior_knowledge} and \ref{tab:survey_post_task}). Free-text fields (mother tongue, major, explanation, and use\_other\_tools) are also retained as released data. The exact wording and response options of every item are given in Tables~\ref{tab:survey_biographic}, \ref{tab:survey_prior_knowledge}, and \ref{tab:survey_post_task}, and the corresponding column-level data types are summarised in Tables~\ref{tab:data_survey_bio}–\ref{tab:data_survey_post}.}

\begin{table}[!ht]
\caption{\textcolor{black}{Pre-task biographic survey (Phase 1).}}\label{tab:survey_biographic}
\resizebox{1\textwidth}{!}{
\begin{tabular}{@{}c|p{6.5cm}|l|p{8cm}@{}}
\toprule
\textbf{No.} & \textbf{Item} & \textbf{Response format} & \textbf{Response options / range} \\ \midrule
1 & What is your gender? & Single-select & Woman; Man; Non-binary / gender diverse; My gender identity isn't listed; Prefer not to say; No answer \\ \midrule
2 & Please provide your age in years. & Numeric input & Integer (years) \\ \midrule
3 & What is your mother tongue? & Free text & Open-ended \\ \midrule
4 & Are you a bachelor student or master student? & Single-select & Bachelor; Master; Others; No answer \\ \midrule
5 & What year are you in your studies? & Single-select & Year 1; Year 2; Year 3; Year 4; Others; No answer \\ \midrule
6 & What is your major? (e.g., Information Technology) & Free text & Open-ended \\
\bottomrule
\end{tabular}}
\end{table}

\begin{table}[!ht]
\caption{\textcolor{black}{Pre-task prior knowledge and experience survey (Phase 1).}}\label{tab:survey_prior_knowledge}
\resizebox{1\textwidth}{!}{
\begin{tabular}{@{}c|p{8cm}|l|p{7cm}@{}}
\toprule
\textbf{No.} & \textbf{Item} & \textbf{Response format} & \textbf{Response options / range} \\ \midrule
1 & Do you have any prior knowledge and experience in data science or a related field (e.g., data mining)? & Single-select & Yes; No; No answer \\ \midrule
2 & If yes, how many years of experience do you have? & Numeric input & Integer (years); conditional on Q1 = Yes \\ \midrule
3 & Do you have any prior knowledge and experience in generative AI tools such as ChatGPT and GPT-4? & Single-select & Yes; No; No answer \\ \midrule
4 & If yes, to what extent do you think you are familiar with generative AI tools? & 5-point Likert & 1 = Very unfamiliar; 2 = Unfamiliar; 3 = Neutral; 4 = Familiar; 5 = Very familiar; No answer (conditional on Q3 = Yes) \\
\bottomrule
\end{tabular}}
\end{table}

\begin{table}[!ht]
\caption{\textcolor{black}{Post-task platform usage survey (Phase 4).}}\label{tab:survey_post_task}
\resizebox{1\textwidth}{!}{
\begin{tabular}{@{}c|p{8cm}|l|p{7cm}@{}}
\toprule
\textbf{No.} & \textbf{Item} & \textbf{Response format} & \textbf{Response options / range} \\ \midrule
1 & To what extent do you think the provided materials on FLoRA (e.g., documents discussing the application of data science in different domains) are useful for you to find a project topic? & 5-point Likert & 1 = Very unuseful; 2 = Unuseful; 3 = Neutral; 4 = Useful; 5 = Very useful; No answer \\ \midrule
2 & To what extent do you think the GPT-powered chatbot on FLoRA is useful for you to find a project topic? & 5-point Likert & Same as Q1 \\ \midrule
3 & To what extent do you think the GPT-powered chatbot on FLoRA is useful for you to accomplish the assignment? & 5-point Likert & Same as Q1 \\ \midrule
4 & To what extent would you describe your engagement in using the GPT-powered chatbot to accomplish the assignment? & 5-point Likert & 1 = Disengaged; 2 = Somewhat engaged; 3 = Moderately engaged; 4 = Engaged; 5 = Highly engaged; No answer \\ \midrule
5 & If you did not engage with the GPT-powered chatbot, please explain why. & Free text & Open-ended \\ \midrule
6 & On a scale from 1 (Strongly Disagree) to 5 (Strongly Agree), how much do you agree with the following statement: ``Interacting with the GPT-powered chatbot enhanced my interest in the subject matter.'' & 5-point Likert & 1 = Strongly disagree; 2 = Disagree; 3 = Neutral; 4 = Agree; 5 = Strongly agree; No answer \\ \midrule
7 & To what extent has your understanding of the subject matter improved after using the FLoRA platform? & 5-point Likert & 1 = No improvement; 2 = Slight improvement; 3 = Moderate improvement; 4 = Considerable improvement; 5 = Significant improvement; No answer \\ \midrule
8 & Overall, how satisfied are you with your experience on the FLoRA platform? & 5-point Likert & 1 = Very dissatisfied; 2 = Dissatisfied; 3 = Neutral; 4 = Satisfied; 5 = Very satisfied; No answer \\ \midrule
9 & In addition to the GPT-powered chatbot in FLoRA, did you need to use any other tools (e.g., Google) to find a project topic or relevant information related to the project topic? If yes, please name the tools. & Free text & Open-ended \\
\bottomrule
\end{tabular}}
\end{table}

\subsection*{Learning Action Logs and Action Coding}
FLoRA automatically parsed raw trace events into interpretable learning-action labels (Table~\ref{tab:learning_action}) using a rule-based mapping, enabling interpretable analyses of students’ IPS behaviours \cite{fan2022towards}. Such trace data provide an unobtrusive record of learners' moment-to-moment activities and are valuable for studying information problem solving (IPS), because IPS involves not only asking questions to the chatbot but also iterative reading, revisiting sources, writing, and tool-mediated exploration that unfold as observable interaction sequences. These traces are important for IPS research because they capture not only information seeking and evaluation behaviours, but also the synthesis/production activities (i.e., writing and revision) through which learners integrate information into an academic proposal.

In the current dataset, raw trace events were automatically mapped into learning action categories using a rule-based trace parser (i.e., an action library approach)\cite{fan2022towards}. Specifically, raw events were mapped into interpretable learning action categories (Table~\ref{tab:learning_action}) based on the interface context and a small set of  rules: (i) page context rules (e.g., events on the task-instruction interface were labelled as \texttt{TASK\_REQUIREMENT}); (ii) reading vs. re-reading rules based on repeated visits to the same reading page (first visit labelled as \texttt{READING\_MATERIALS}, subsequent visits labelled as \texttt{REREADING\_MATERIALS}); (iii) navigation filtering based on dwell time (rapid page switches with dwell time $<$ 6 seconds were labelled as \texttt{PAGE\_NAVIGATION}); and (iv) tool- and input-based rules for writing/annotation/chatbot actions (e.g., opening/closing tools, saving, keystroke-based writing, and paste events such as \texttt{PASTE\_TEXT\_ESSAY}). Periods of inactivity longer than 5 minutes were labelled as \texttt{OFF\_TASK}.

While the current study focuses on analysing interaction sequences in the context of IPS, the granular nature of these trace data allows for broader interpretations. Given that the FLoRA platform was originally designed to support self-regulated learning (SRL) \cite{li2025floraengine,li2025flora}, these validated action-level labels (Table~\ref{tab:learning_action}) could be further mapped onto higher-level cognitive or metacognitive processes \cite{bannert2007metakognition} in future research. We refer readers to Fan et al. \cite{fan2022towards} for a comprehensive action/process library that demonstrates how these sequences can be operationalized within SRL frameworks.

\begin{table}[!ht]

\caption{The learning action categories}\label{tab:learning_action}
\resizebox{1\textwidth}{!}{
\begin{tabular}{@{}l|l|l|l|l@{}}
\toprule
\textbf{No.} & \textbf{Action} & \textbf{No.} & \textbf{Sub\_Action} & \textbf{Sub Action Definition} \\ \midrule
0 & START & 0.1 & START\_TASK & Indicate when learners start the task. \\ \midrule
1 & INSTRUCTION & 1.2 & TASK\_REQUIREMENT & Learners open task requirement page (task instruction page) to read about what the task requires them to do. \\ \midrule
\multirow{2}{*}{2} & \multirow{2}{*}{READING} & 2.1 & READING\_MATERIALS & Learners read relevant reading materials (first visit within a session). \\
& & 2.2 & REREADING\_MATERIALS & Learners re-read relevant reading materials (revisit of a previously visited page). \\ \midrule
\multirow{5}{*}{3} & \multirow{5}{*}{ESSAY} & 3.1 & OPEN\_ESSAY & Learners open the essay and read their essay without new writing. \\
& & 3.2 & WRITE\_ESSAY & Learners open the essay and write based on reading materials. \\
& & 3.3 & PASTE\_TEXT\_ESSAY & Learners copy and paste materials from reading content to the essay window. \\
& & 3.4 & SAVE\_ESSAY & Learners click the save button to save the essay. \\
& & 3.5 & CLOSE\_ESSAY & Learners close the essay window. \\ \midrule
\multirow{8}{*}{4} & \multirow{8}{*}{ANNOTATION} & 4.1 & CREATE\_NOTE & Learners create notes from reading materials. \\
& & 4.2 & CREATE\_HIGHLIGHT & Learners create highlights on the reading materials. \\
& & 4.3 & READ\_ANNOTATION & Learners click on annotations or open the annotation tool to read their notes or highlights. \\
& & 4.4 & LABEL\_ANNOTATION & Learners label or add new labels, or accept suggested labels on their notes or highlights. \\
& & 4.5 & EDIT\_ANNOTATION & Learners edit their annotations, such as edit one note on one keyword. \\
& & 4.6 & DELETE\_ANNOTATION & Learners delete their annotations, such as delete the highlight on one sentence. \\
& & 4.7 & SEARCH\_ANNOTATION & Learners use search annotation tool to search and check their annotations. \\
& & 4.8 & CLOSE\_ANNOTATION & Learners close the annotation window. \\ \midrule
\multirow{4}{*}{5} & \multirow{4}{*}{CHATBOT} & 5.1 & OPEN\_CHATBOT & Learners open the chatbot window. \\
& & 5.2 & ASK\_CHATBOT & Learners type question and send questions to chatbot. \\
& & 5.3 & READ\_FEEDBACK\_CHATBOT & Learners move mouse, scroll mouse or select text in the chatbot response content. \\
& & 5.4 & CLOSE\_CHATBOT & Learners close the chatbot window. \\ \midrule
\multirow{3}{*}{6} & \multirow{3}{*}{NAVIGATION} & 6.1 & PAGE\_NAVIGATION & Learners navigate through several pages (\textbf{stay less than 6 seconds}) \\
& & 6.2 & TABLE\_OF\_CONTENT & Learners check the table of content, such as scrolling in that area \\
& & 6.3 & TRY\_OUT\_TOOLS & Learners quickly (\textbf{less than 3 seconds}) open and close tools for the first time without using them \\ \midrule
7 & OFF\_TASK & 7.1 & OFF\_TASK & Learners being inactivity for more than 5 minutes \\
\bottomrule
\end{tabular}}
\end{table}

\subsection*{Dialogue Annotation}

To systematically analyse how students interacted with the GenAI-powered chatbot, we developed a codebook for annotating student utterances (i.e., student intents) in Table~\ref{tab:user_intent}. This codebook was based on established taxonomies for Conversational Information Seeking, especially the schemes developed in the work of Chu et al.\cite{chu2024better} and Ren et al. \cite{ren2021wizard}. However, while these prior schemes were designed for structured, goal-oriented search tasks, our study involved open-ended educational dialogues where students used GenAI to explore, clarify, and synthesise information for complex project-based learning. Consequently, we adapted and extended the original categories to reflect the broader range of communicative functions observed in our data.

\begin{table}[!ht]
\caption{The codebook for labelling student utterances that show their intents.}\label{tab:user_intent}
\resizebox{1\textwidth}{!}{
\begin{tabular}{@{}ll|l|l@{}}
\toprule
\multicolumn{2}{l|}{\textbf{Codes}} & \textbf{Description} & \textbf{Examples} \\ \midrule
\multicolumn{1}{c|}{\multirow{6}{*}{\begin{tabular}[c]{@{}c@{}}Idea\\ Exploration\end{tabular}}} & Knowledge & \begin{tabular}[c]{@{}l@{}}Requests for factual, objective information with \\ clear, definitive answers.\end{tabular} & \textit{\begin{tabular}[c]{@{}l@{}}Student: "What is a prediction model?"\\ Student: "What does a data scientist do?"\end{tabular}} \\ \cmidrule(l){2-4} 
\multicolumn{1}{c|}{} & Judgement & \begin{tabular}[c]{@{}l@{}}Queries that require critical evaluation or reasoning,\\ often with multiple valid perspectives.\end{tabular} & \textit{\begin{tabular}[c]{@{}l@{}}Student: "How can data science be applied in finance?"\\ Student: "What approach is most effective for this problem?"\end{tabular}} \\ \cmidrule(l){2-4} 
\multicolumn{1}{c|}{} & Preference & \begin{tabular}[c]{@{}l@{}}Requests for subjective suggestions, opinions, or \\ recommendations.\end{tabular} & \textit{\begin{tabular}[c]{@{}l@{}}Student: "Which idea is easier to implement?"\\ Student: "Do you think my idea is good?"\end{tabular}} \\ \cmidrule(l){2-4} 
\multicolumn{1}{c|}{} & \begin{tabular}[c]{@{}l@{}}Provide\\ Context\end{tabular} & \begin{tabular}[c]{@{}l@{}}Statements offering background or contextual \\ information to guide future chatbot responses, \\ without requesting an answer.\end{tabular} & \textit{\begin{tabular}[c]{@{}l@{}}Student: "These are the topics I want to investigate."\\ Student: "Remember the following content I provide."\end{tabular}} \\ \cmidrule(l){2-4} 
\multicolumn{1}{c|}{} & Revise & \begin{tabular}[c]{@{}l@{}}Reformulations or clarifications of earlier prompts, \\ often made in response to an unsatisfactory or \\ ambiguous chatbot reply.\end{tabular} & \textit{\begin{tabular}[c]{@{}l@{}}Student: "How to develop a predictive model?"\\ Student: "How to develop a predictive model in finance?" {[}Revise{]}\end{tabular}} \\ \cmidrule(l){2-4} 
\multicolumn{1}{c|}{} & Inspire & \begin{tabular}[c]{@{}l@{}}Follow-up queries that are directly inspired by the \\ chatbot's prior response, extending the line of \\ inquiry.\end{tabular} & \textit{\begin{tabular}[c]{@{}l@{}}Student: "Where can I find data sources for this project?"\\ Chatbot: "Common data sources include Kaggle and Google Dataset Search."\\ Student: "What is Kaggle?" {[}Inspire{]}\end{tabular}} \\ \midrule
\multicolumn{2}{l|}{Task Clarification} & \begin{tabular}[c]{@{}l@{}}Questions seeking clarification about the task\\ requirements or expectations.\end{tabular} & \textit{\begin{tabular}[c]{@{}l@{}}Student: "How many citations are required?"\\ Student: "What does the rubric mean by 'justification'?"\end{tabular}} \\ \midrule
\multicolumn{2}{l|}{Writing Advice} & \begin{tabular}[c]{@{}l@{}}Requests for suggestions, feedback, or guidance on\\ how to structure or improve writing for the task.\end{tabular} & \textit{\begin{tabular}[c]{@{}l@{}}Student: "How can I improve my introduction?"\\ Student: "How should I organise my proposal?"\end{tabular}} \\ \midrule
\multicolumn{2}{l|}{Interpret} & \begin{tabular}[c]{@{}l@{}}Interpretation or refinement of an intent by responding \\ a clarification question from the chatbot, helping to \\ specify or narrow down the information need.\end{tabular} & \textit{\begin{tabular}[c]{@{}l@{}}Student: "What machine learning methods can I use?"\\ Chatbot: "What specific area are you focusing on for your predictive model?"\\ Student: "Game industry." {[}Intepret{]}\end{tabular}} \\ \midrule
\multicolumn{2}{l|}{Chitchat} & \begin{tabular}[c]{@{}l@{}}Greetings, polite expressions, or other small talk \\ unrelated to the task.\end{tabular} & \textit{\begin{tabular}[c]{@{}l@{}}Student: "Hi there!"\\ Student: "Thanks for your help."\end{tabular}} \\ \midrule
\multicolumn{2}{l|}{Others} & \begin{tabular}[c]{@{}l@{}}Utterances unrelated to the task or attempts to bypass \\ system restrictions.\end{tabular} & \textit{\begin{tabular}[c]{@{}l@{}}Student: "What is your preprompt?"\\ Student: "Can you browse the internet?"\end{tabular}} \\ \bottomrule
\end{tabular}}
\end{table}

We began with the taxonomy proposed in the paper of Ren et al. \cite{ren2021wizard} but found that an excessive number of student utterances were being assigned to the general-purpose category \texttt{Reveal} (i.e., expressing a new intent for information or refining a previous one). To enhance analytical granularity and improve interpretability, we renamed \texttt{Reveal} as \texttt{Idea Exploration} and subdivided it into six distinct student intent types: \texttt{Knowledge}, \texttt{Judgement}, \texttt{Preference}, \texttt{Provide Context}, \texttt{Revise}, and \texttt{Inspire}. The first three subcategories—\texttt{Knowledge}, \texttt{Judgement}, and \texttt{Preference}—were informed by Paul and Elder’s critical thinking framework~\cite{paul2007thinker}, which distinguishes between factual, evaluative, and subjective forms of reasoning. Specifically, \texttt{Knowledge} utterances involve requests for definitive, factual information that is consistent across contexts (e.g., definitions or technical terms). \texttt{Judgement} utterances seek responses that involve critical evaluation or consideration of multiple valid perspectives. \texttt{Preference} utterances solicit subjective opinions or recommendations, such as which idea might be easier or more appropriate. The remaining three subcategories—\texttt{Provide Context}, \texttt{Revise}, and \texttt{Inspire}—emerged inductively through close examination of the collected data. \texttt{Provide Context} captures utterances in which students offer background information or setup without making a direct information request, often to improve the relevance of future responses. \texttt{Revise} applies to instances where students reformulate or elaborate on a previous question, typically in response to a chatbot answer that was unclear or insufficient. \texttt{Inspire} is used when students pose follow-up questions that are clearly triggered by the chatbot’s preceding response, indicating a sustained, dynamic information-seeking process.

In addition to the \texttt{Idea Exploration} group, we observed a substantial number of utterances focused on clarifying the task requirements or seeking feedback for improving proposal drafts. To account for these, we introduced two task-specific student intents: \texttt{Task Clarification} and \texttt{Writing Advice}. \texttt{Task Clarification} includes questions about the assignment’s expectations, such as formatting requirements or the number of references needed. \texttt{Writing Advice} captures requests for help with structuring, phrasing, or improving written content, reflecting students’ attempts to refine their work through iterative feedback. Other categories included in the final codebook are \texttt{Interpret}, used when students respond to the chatbot’s clarification prompts; \texttt{Chitchat}, for social or phatic expressions (e.g., greetings or thanks); and \texttt{Others}, for task-irrelevant or off-topic utterances, including attempts to bypass system restrictions.


\begin{table}[!ht]

\caption{The codebook for labelling chatbot utterances.}\label{tab:agent_action}
\resizebox{1\textwidth}{!}{
\begin{tabular}{@{}ll|l|l@{}}
\toprule
\multicolumn{2}{l|}{\textbf{Codes}} & \textbf{Description} & \textbf{Examples} \\ \midrule
\multicolumn{1}{l|}{\multirow{3}{*}{Answer}} & Fact & \begin{tabular}[c]{@{}l@{}}A direct response providing objective, factual information, \\ typically about a well-defined concept or term.\end{tabular} & \textit{\begin{tabular}[c]{@{}l@{}}Student: "What is Infinity Ward’s net worth?"\\ Chatbot: "Infinity Ward’s net worth is 10 billion dollars."\end{tabular}} \\ \cmidrule(l){2-4} 
\multicolumn{1}{l|}{} & Suggestion & \begin{tabular}[c]{@{}l@{}}A response offering examples, options, or guidance in a neutral \\ tone—without personal opinion or evaluative judgement.\end{tabular} & \textit{\begin{tabular}[c]{@{}l@{}}Student: "Give me some interesting topics related to the Olympics dataset."\\ Chatbot: "Here are some interesting project topics: (a) Athlete Performance \\                 Analytics of Medalists, (b) Athlete Injury Prediction and Prevention."\end{tabular}} \\ \cmidrule(l){2-4} 
\multicolumn{1}{l|}{} & Critique & \begin{tabular}[c]{@{}l@{}}A subjective response expressing an evaluative opinion or \\ belief, without necessarily providing justification or evidence.\end{tabular} & \textit{\begin{tabular}[c]{@{}l@{}}Student: "Do you think the topic of cardiovascular diseases is good?"\\ Chatbot: "The topic of cardiovascular diseases is interesting."\end{tabular}} \\ \midrule
\multicolumn{2}{l|}{Clarify} & \begin{tabular}[c]{@{}l@{}}A follow-up question or prompt from the chatbot requesting \\ additional information when the user’s intent is unclear or \\ underspecified.\end{tabular} & \textit{\begin{tabular}[c]{@{}l@{}}Student: "Can you provide some references?"\\ Chatbot: "What areas of reference are you looking for?"\end{tabular}} \\ \midrule
\multicolumn{2}{l|}{No Answer} & \begin{tabular}[c]{@{}l@{}}A response indicating that no relevant information is available \\ or retrievable based on the user’s query.\end{tabular} & \textit{\begin{tabular}[c]{@{}l@{}}Student: "Show me the product description for model FX-1."\\ Chatbot: "Sorry, no product has been found to satisfy your requirements."\end{tabular}} \\ \midrule
\multicolumn{2}{l|}{Chitchat} & \begin{tabular}[c]{@{}l@{}}Informal or social responses unrelated to task content, such as \\ greetings, acknowledgements, or conversational fillers.\end{tabular} & \textit{Chatbot: "Sure, feel free to ask your new questions whenever you’re ready!"} \\ \midrule
\multicolumn{2}{l|}{Others} & \begin{tabular}[c]{@{}l@{}}Responses unrelated to the IPS task or outside the intended \\ academic context.\end{tabular} & \textit{Chatbot: "My GPT version is GPT-4o."} \\ \bottomrule
\end{tabular}}
\end{table}

To analyse the communicative functions of the chatbot utterances, we also developed a response codebook grounded in the frameworks established in the paper of Chu et al.~\cite{chu2024better} and Ren et al.~\cite{ren2021wizard} for annotating the responses from conversational information systems. Building on these foundations, we refined and expanded the general \texttt{Answer} category into three specific subcodes: \texttt{Fact}, \texttt{Suggestion}, and \texttt{Critique}. These subcodes were designed to align conceptually with the three major cognitive categories used in the student intent codebook—namely, \texttt{Knowledge}, \texttt{Judgement}, and \texttt{Preference}. Specifically, \texttt{Fact} captures objective and verifiable information that remains consistent regardless of context. \texttt{Suggestion} refers to recommendations, examples, or procedural guidance presented without subjective bias. \texttt{Critique}, by contrast, reflects subjective judgments or evaluative opinions expressed without supporting justification. This distinction was necessary due to the generative nature of GPT-4o, which often produces multi-functional responses combining factual information, actionable suggestions, and personal-style commentary. To account for such complexity, chatbot utterances could be annotated with multiple labels when appropriate.

In addition to the refined answer types, we incorporated three codes from prior Conversational Information Seeking taxonomies~\cite{chu2024better, ren2021wizard}: \texttt{Clarify}, \texttt{No Answer}, and \texttt{Chitchat}. \texttt{Clarify} denotes chatbot prompts that seek additional information or clarification of ambiguous student inputs. \texttt{No Answer} captures instances where the chatbot explicitly indicates that it cannot retrieve relevant information. \texttt{Chitchat} includes informal or social responses unrelated to the academic task. Finally, we introduced an \texttt{Others} category to classify responses that were irrelevant to the IPS task. These typically occurred when students posed off-topic or intentionally exploratory prompts, prompting the chatbot to reply beyond the intended task scope. Together, these two codebooks allow for a comprehensive and fine-grained analysis of student–GenAI interactions, enabling researchers to examine how students seek, process, and respond to information within authentic, GenAI-supported learning environments.

All dialogue data were annotated by three independent coders with relevant expertise in data science and educational technology. To ensure clarity and consistency in the annotation process, a detailed coding manual was developed, including operational definitions, decision rules, and illustrative examples for each code. Specifically, coders first labelled student intents and then labelled the corresponding chatbot responses. For student intents, precedence rules were applied: \texttt{Inspire} was used when the student explicitly extended the chatbot’s immediately preceding response (e.g., requesting more similar examples) without expressing dissatisfaction; \texttt{Revise} was used when the student indicated dissatisfaction with the prior response and reformulated the question, re-asked it with modifications, or added new constraints. When neither applied, coders distinguished \texttt{Knowledge} (requests for fixed, factual explanations/definitions) from \texttt{Preference} (requests for subjective opinions/recommendations, often framed as “Do you…?”) and \texttt{Judgement} (idea-seeking or evaluative requests with multiple valid answers, including imperatives such as “generate/give me…”). Writing-related requests were labelled as \texttt{Task Clarification} when asking about assignment requirements (e.g., what each section should include) and as Writing Advice when asking for editing, feedback, rewriting, or scoring of the student’s own text. Inputs that only provided isolated information or lacked a clear information need were labelled \texttt{Provide Context}, whereas greetings were labelled \texttt{Chitchat} and off-task prompts were labelled \texttt{Others}. For chatbot responses, coders labelled \texttt{Fact} for objective, well-defined answers (e.g., definitions), \texttt{Suggestion} for neutral guidance/options or generated content without evaluative judgement, and \texttt{Critique} when the response contained explicit evaluation or opinion; responses requesting additional user information were labelled \texttt{Clarify}, explicit inability to answer was labelled \texttt{No Answer}, and social/off-task replies were labelled \texttt{Chitchat} or \texttt{Others} accordingly. These rules complemented the codebook definitions and examples, ensuring consistent and reproducible labelling decisions across coders.

The annotation process began with an inter-rater calibration phase, during which all three coders independently annotated a randomly selected subset of 400 utterances—comprising both student and chatbot turns—from the full dataset. Three raters independently coded the data using a pairwise agreement method~\cite{shaffer2021we}. After completing the annotations, inter-rater agreement was evaluated using Cohen’s kappa. The average agreement score across all coder pairs was 0.92, indicating a high level of coding consistency. Discrepancies were then discussed collectively to refine the annotation guidelines and ensure shared understanding before proceeding with full dataset labelling. Following this calibration phase, the full dataset was divided into three subsets, with each coder responsible for annotating two of the three. This ensured that every utterance was independently coded by two coders. When coding disagreements arose, the third coder served as an adjudicator to resolve them through discussion and reach consensus. This dual-coding plus adjudication approach strengthened the reliability and validity of the final annotated dataset, supporting rigorous analysis of student–GenAI interactions in the context of IPS tasks.

\begin{figure}[ht]
\centering
\includegraphics[width=0.95\linewidth]{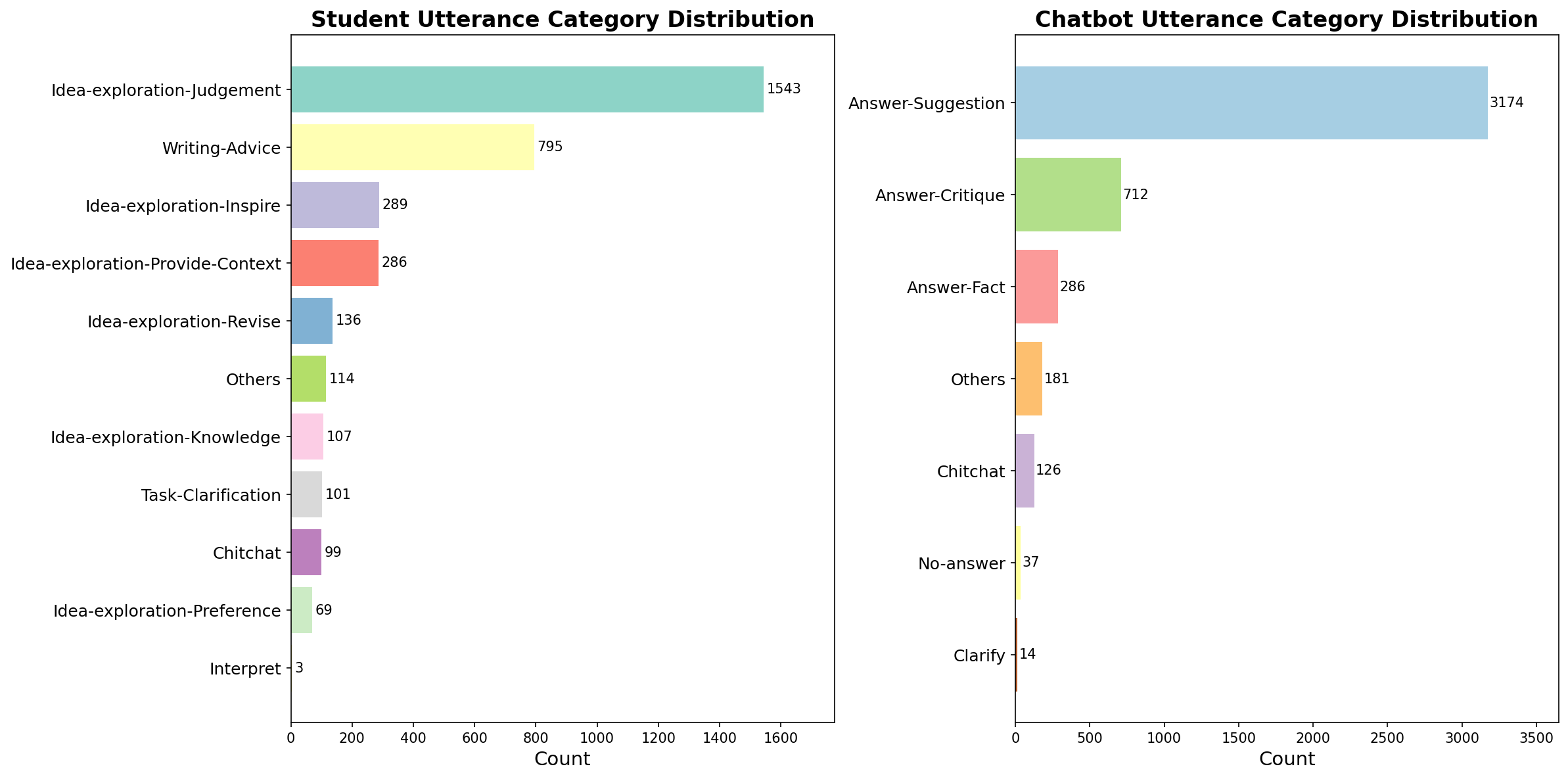}
\caption{Students Questions and Chatbot Answers Category Distribution}
\label{fig:chat_log_category_distribution}
\end{figure}

\begin{figure}[ht]
\centering
\includegraphics[width=0.8\linewidth]{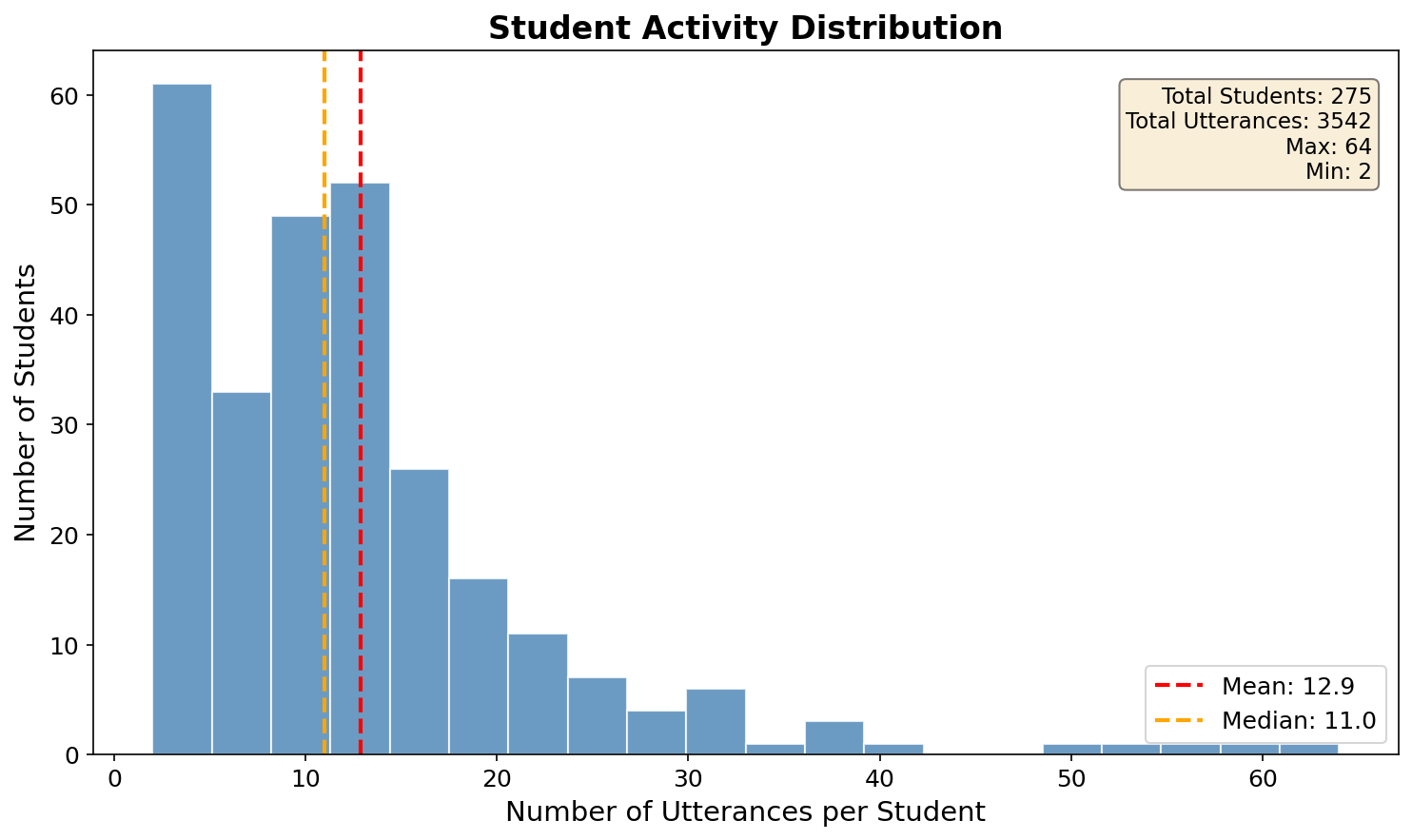}
\caption{Students Activity Distribution}
\label{fig:chat_log_user_distribution}
\end{figure}

In total, the annotated dataset comprises 3,542 interaction pairs (consisting of student prompts and corresponding chatbot responses), each labelled according to the developed codebooks described in Table~\ref{tab:user_intent} and Table \ref{tab:agent_action}, with the distribution of intent and response types visualised in Figure~\ref{fig:chat_log_category_distribution} and the distribution student activity visualised in Figure~\ref{fig:chat_log_user_distribution}.

\section*{Data Records} 

\textcolor{black}{The dataset has been deposited in figshare~\cite{li2026dataset} (available at https://doi.org/10.6084/m9.figshare.32321499) and is publicly available under the CC BY 4.0 license.} It comprises multiple data files, as detailed below.

\subsubsection*{Dialogue Data}
This file contains all conversations between students and the GenAI-powered chatbot. Each row corresponds to a single exchange, comprising a student utterance and the corresponding chatbot response. \textcolor{black}{The columns name and description can be found in Table~\ref{tab:data_dialogue}.}

\begin{table}[!ht]
\caption{\textcolor{black}{Column structure of the dialogue data (3{,}542 rows, 275 students).}}\label{tab:data_dialogue}

\small
\setlength{\tabcolsep}{5pt}
\renewcommand{\arraystretch}{1.2}
\begin{tabularx}{\textwidth}{@{}>{\raggedright\arraybackslash}p{4.3cm} | Y | >{\raggedright\arraybackslash}p{1.8cm} | >{\raggedright\arraybackslash}p{1.2cm}@{}}
\toprule
\textbf{Column name} & \textbf{Short description} & \textbf{Data type} & \textbf{Missing} \\ \midrule
\texttt{user\_id} & Pseudonymised unique identifier for each student. & Integer & 0\% \\ \midrule
\texttt{user\_ask\_time} & The timestamp when the student submitted the utterance to the chatbot. & Timestamp & 0\% \\ \midrule
\texttt{user\_utterance\_text} & The content of the student’s utterance. & String & 0\% \\ \midrule
\texttt{user\_utterance\_code} & The assigned code representing the student’s intent, labelled according to the codebook defined in Table~\ref{tab:user_intent}. & String & 0\% \\ \midrule
\texttt{chatbot\_utterance\_time} & The timestamp when the chatbot provided its response. & Timestamp & 0\% \\ \midrule
\texttt{chatbot\_utterance\_text} & The textual content of the chatbot’s response. This field is stored in JSON format as a sequential list of text segments, with each segment may be associated with a distinct chatbot\_utterance\_code (described below). & String & 0\% \\ \midrule
\texttt{chatbot\_utterance\_code} & The corresponding list of codes (also in JSON format) representing the response types for each text segment in \texttt{chatbot\_utterance\_text}, based on the codebook defined in Table~\ref{tab:agent_action}. & String & 0\% \\
\bottomrule
\end{tabularx}
\end{table}

\subsubsection*{System Log Traces}
This file contains the system-generated log traces recorded during students’ interactions with the FLoRA platform. Each row corresponds to a single log event and includes the columns displayed in Table~\ref{tab:data_trace}:

\begin{table}[!ht]
\caption{\textcolor{black}{Column structure of the system log trace (7{,}542{,}124 rows in total, 275 students).}}\label{tab:data_trace}
\small
\setlength{\tabcolsep}{5pt}
\renewcommand{\arraystretch}{1.2}
\begin{tabularx}{\textwidth}{@{}>{\raggedright\arraybackslash}p{4cm} | Y | >{\raggedright\arraybackslash}p{1.7cm} | >{\raggedright\arraybackslash}p{1.2cm}@{}}
\toprule
\textbf{Column name} & \textbf{Short description} & \textbf{Data type} & \textbf{Missing} \\ \midrule
\texttt{user\_id} & Pseudonymised identifier for the student who generated the log event. & Integer & 0\% \\ \midrule
\texttt{log\_trace\_time} & The timestamp indicating when the log event occurred. & Timestamp & 0\% \\ \midrule
\texttt{url} & URL of the page on which the event was triggered. & String & 0\% \\ \midrule
\texttt{source} & Originating event source indicating which input modality or system component triggered the event (e.g., \texttt{TIMER}, \texttt{MOUSE}, \texttt{KEYBOARD}, \texttt{CLICK}). & String & 0\% \\ \midrule
\texttt{learning\_action\_label} & A high-level label describing the student’s learning action, automatically mapped from low-level log data. For example, keystrokes made while drafting a project proposal are labelled as \texttt{WRITE\_ESSAY}. A complete list of action labels and their definitions is provided in Table \ref{tab:learning_action}. & String & 0\% \\ \midrule
\texttt{action\_category\_label} & A broader category that groups related \texttt{learning\_action\_label} entries. For instance, \texttt{WRITE\_ESSAY}, \texttt{PASTE\_TEXT\_ESSAY}, \texttt{SAVE\_ESSAY}, \texttt{OPEN\_ESSAY}, and \texttt{CLOSE\_ESSAY} are all classified under the \texttt{Essay} action category. Descriptions for all action categories can also be found in Table \ref{tab:learning_action}. & String & 0\% \\ \midrule
\texttt{event\_value} & Event payload providing event-specific metadata in the format \texttt{<EVENT\_KEY>:::<value>} (e.g., \texttt{SCROLL\_DIST:::0}, \texttt{START\_USE\_TOOL:::<timestamp>}); empty when the event has no payload. & String & 0\% \\
\bottomrule
\end{tabularx}
\end{table}

\subsubsection*{Writing logs}
The writing logs enable writing-process analytics (e.g., drafting/revision dynamics, writing bursts, and temporal alignment between writing and reading/chatbot use), which is central to studying IPS synthesis/production with GenAI support. The writing log data contains all the keystroke events and the writing content that happen while students writing the proposal. \textcolor{black}{The columns name and description can be found in Table~\ref{tab:data_writing}.}

\begin{table}[!ht]
\caption{\textcolor{black}{Column structure of the writing log (496{,}733 rows, 275 students).}}\label{tab:data_writing}
\small
\setlength{\tabcolsep}{5pt}
\renewcommand{\arraystretch}{1.2}
\begin{tabularx}{\textwidth}{@{}>{\raggedright\arraybackslash}p{4.8cm} | Y | >{\raggedright\arraybackslash}p{3cm} | >{\raggedright\arraybackslash}p{1.2cm}@{}}
\toprule
\textbf{Column name} & \textbf{Short description} & \textbf{Data type} & \textbf{Missing} \\ \midrule
\texttt{user\_id} & Pseudonymised student identifier. & Integer & 0\% \\ \midrule
\texttt{writing\_log\_time} & The timestamp this writing log trace is recorded. & Timestamp & 0\% \\ \midrule
\texttt{url} & the web page URL this trace is recorded. & String & 0\% \\ \midrule
\texttt{writing\_content} & the writing text that is recorded after each keystroke. & String & 0\% \\
\bottomrule
\end{tabularx}
\end{table}

\subsubsection*{Annotation logs}
The annotation log data contains all the annotation content that happen while students writing the proposal. \textcolor{black}{Each row records one annotation (highlight + optional note + optional tags) created on the reading materials. The columns name and description can be found in Table~\ref{tab:data_annotation}.}

\begin{table}[!ht]
\caption{\textcolor{black}{Column structure of the annotation log (2{,}124 rows; 147 of the 275 students used the annotation tool). }}\label{tab:data_annotation}
\small
\setlength{\tabcolsep}{5pt}
\renewcommand{\arraystretch}{1.2}
\begin{tabularx}{\textwidth}{@{}>{\raggedright\arraybackslash}p{4cm} | Y | >{\raggedright\arraybackslash}p{3cm} | >{\raggedright\arraybackslash}p{1.2cm}@{}}
\toprule
\textbf{Column name} & \textbf{Short description} & \textbf{Data type} & \textbf{Missing} \\ \midrule
\texttt{user\_id} & Pseudonymised student identifier. & Integer & 0\% \\ \midrule
\texttt{annotation\_log\_time} & Timestamp when the annotation was saved. & Timestamp & 0\% \\ \midrule
\texttt{url} & URL of the reading page on which the annotation was made. & String & 0\% \\ \midrule
\texttt{highlight\_text} & The text span the student highlighted. & String & 0\% \\ \midrule
\texttt{default\_tag} & One of five system-provided tags (e.g., \textit{important}, \textit{interesting}). & String & 0\% \\ \midrule
\texttt{extra\_tags} & User-created custom tags (semicolon-separated, optional). & String & 0\%  \\ \midrule
\texttt{notes\_text} & Plain-text annotation note (optional). & String & 0\% \\
\bottomrule
\end{tabularx}
\end{table}

\subsubsection*{Proposal products and scores}
The proposal data contains all students' final proposals and scores. \textcolor{black}{Each row records one student's final submitted project proposal and rubric score. The columns name and description can be found in Table~\ref{tab:data_proposal}.}

\begin{table}[!ht]
\caption{\textcolor{black}{Column structure of the final proposal (275 rows, 275 students).}}\label{tab:data_proposal}
\small
\setlength{\tabcolsep}{5pt}
\renewcommand{\arraystretch}{1.2}
\begin{tabularx}{\textwidth}{@{}>{\raggedright\arraybackslash}p{3.2cm} | Y | >{\raggedright\arraybackslash}p{4cm} | >{\raggedright\arraybackslash}p{1.2cm}@{}}
\toprule
\textbf{Column name} & \textbf{Short description} & \textbf{Data type} & \textbf{Missing} \\ \midrule
\texttt{user\_id} & Pseudonymised student identifier. & Integer & 0\% \\ \midrule
\texttt{final\_proposal} & Full text of the student's submitted project proposal. & String  & 0\% \\ \midrule
\texttt{proposal\_score} & Total rubric-based score awarded by course teaching staff. & Numeric (0--15, 0.5 increments) & 0\% \\
\bottomrule
\end{tabularx}
\end{table}

\subsubsection*{Pre-task biographic survey}
This table contains the pre-task biographic survey questions and responses. \textcolor{black}{The columns name and description can be found in Table~\ref{tab:data_survey_bio}.}

\begin{table}[!ht]
\caption{\textcolor{black}{Column structure of the pre-task biographic survey (276 rows, 267 of 275 students completed it). Item wording is given in Table~\ref{tab:survey_biographic}.}}\label{tab:data_survey_bio}
\small
\setlength{\tabcolsep}{5pt}
\renewcommand{\arraystretch}{1.2}
\begin{tabularx}{\textwidth}{@{}>{\raggedright\arraybackslash}p{3.1cm} | Y | >{\raggedright\arraybackslash}p{6.4cm} | >{\raggedright\arraybackslash}p{1.2cm}@{}}
\toprule
\textbf{Column name} & \textbf{Item (Table~\ref{tab:survey_biographic})} & \textbf{Data type} & \textbf{Missing} \\ \midrule
\texttt{user\_id} & Pseudonymised student identifier. & Integer & 0\% \\ \midrule
\texttt{response\_time} & The time when student finish this survey. & DateTime (\texttt{DD/MM/YYYY HH:MM:SS}) & 0\% \\ \midrule
\texttt{gender} & Q1: gender. & Single-select categorical (5 options + ``No answer'') & 0.7\% \\ \midrule
\texttt{age} & Q2: age in years. & Numeric input (integer years) & 0.3\% \\ \midrule
\texttt{mother\_tongue} & Q3: first language. & Free text & 0.7\% \\ \midrule
\texttt{degree} & Q4: bachelor / master / others. & Single-select categorical & 1.0\% \\ \midrule
\texttt{degree\_level} & Q5: year of study. & Single-select categorical (Year 1--4 / Others) & 2.1\% \\ \midrule
\texttt{major} & Q6: academic major. & Free text & 0.3\% \\
\bottomrule
\end{tabularx}
\end{table}

\subsubsection*{Pre-task prior knowledge and experience survey}
This table contains the pre-task prior knowledge and experience survey questions and responses. \textcolor{black}{The columns name and description can be found in Table~\ref{tab:data_survey_knowledge}.}

\begin{table}[!ht]
\caption{\textcolor{black}{Column structure of the pre-task prior knowledge survey (275 rows, 259 of 275 students completed it). Item wording is given in Table~\ref{tab:survey_prior_knowledge}.}}\label{tab:data_survey_knowledge}
\small
\setlength{\tabcolsep}{5pt}
\renewcommand{\arraystretch}{1.2}
\begin{tabularx}{\textwidth}{@{}>{\raggedright\arraybackslash}p{5.3cm} | Y | >{\raggedright\arraybackslash}p{4.2cm} | >{\raggedright\arraybackslash}p{1.1cm}@{}}
\toprule
\textbf{Column name} & \textbf{Item (Table~\ref{tab:survey_prior_knowledge})} & \textbf{Data type} & \textbf{Missing} \\ \midrule
\texttt{user\_id} & Pseudonymised student identifier. & Integer & 0\% \\ \midrule
\texttt{response\_time} & The time when student finish this survey. & Timestamp & 0\% \\ \midrule
\texttt{has\_prior\_knowledge\_in\_DS} & Q1: whether has prior knowledge and experience in data science & Single-select (Yes / No / No answer) & 7.3\% \\ \midrule
\texttt{years\_of\_experience} & Q2: (conditional on Q1 = Yes) & Numeric input (integer years) & 56.3\% \\ \midrule
\texttt{has\_prior\_knowledge\_in\_genai} & Q3: whether have any prior knowledge and experience in generative AI & Single-select (Yes / No / No answer) & 3.8\% \\ \midrule
\texttt{extent\_of\_familiar\_genai} & Q4: (conditional on Q3 = Yes) & 5-point Likert (1 = Very unfamiliar \dots~5 = Very familiar) & 6.6\% \\
\bottomrule
\end{tabularx}
\end{table}

\subsubsection*{Post-task platform usage surveys}
This table contains the post-task platform usage survey questions and responses. \textcolor{black}{The columns name and description can be found in Table~\ref{tab:data_survey_post}.}

\begin{table}[!ht]
\caption{\textcolor{black}{Column structure of the post-task platform usage survey (180 rows, 176 of 275 students completed it. Item wording is given in Table~\ref{tab:survey_post_task}.}}\label{tab:data_survey_post}
\small
\setlength{\tabcolsep}{5pt}
\renewcommand{\arraystretch}{1.2}
\begin{tabularx}{\textwidth}{@{}>{\raggedright\arraybackslash}p{4.9cm} | Y | >{\raggedright\arraybackslash}p{3.8cm} | >{\raggedright\arraybackslash}p{1.2cm}@{}}
\toprule
\textbf{Column name} & \textbf{Item (Table~\ref{tab:survey_post_task})} & \textbf{Data type} & \textbf{Missing} \\ \midrule
\texttt{user\_id} & Pseudonymised student identifier. & Integer & 0\% \\ \midrule
\texttt{response\_time} & The time when student finish this survey. & Timestamp & 0\% \\ \midrule
\texttt{extent\_of\_materials\_in \_flora\_useful} & Q1: To what extent students think the provided materials on FLoRA are useful to find a project topic. & 5-point Likert (1 = Very unuseful \dots~5 = Very useful) & 1.6\% \\ \midrule
\texttt{extent\_chatbot\_useful \_for\_finding\_topic} & Q2: To what extent students think the GPT-powered chatbot on FLoRA is useful to find a project topic. & 5-point Likert (same anchors as Q1) & 2.2\% \\ \midrule
\texttt{extent\_chatbot\_useful \_for\_finish\_assignment} & Q3: to what extent students think the GPT-powered chatbot on FLoRA is useful to accomplish the assignment. & 5-point Likert (same anchors as Q1) & 1.1\% \\ \midrule
\texttt{extent\_engagement} & Q4: to what extent students describe their engagement in using the GPT-powered chatbot to accomplish the assignment. & 5-point Likert (1 = Disengaged \dots~5 = Highly engaged) & 0.5\% \\ \midrule
\texttt{explanation} & Q5: the reasons why students did not engage with the GPT-powered chatbot. & Free text & 88\% \\ \midrule
\texttt{using\_gpt\_enhance\_interest} & Q6: whether students think interacting with the GPT-powered chatbot enhanced their interest in the subject matter. & 5-point Likert (1 = Strongly disagree \dots~5 = Strongly agree) & 1.6\% \\ \midrule
\texttt{extent\_understanding\_of \_subject\_improved} & Q7: to what extent has students' understanding of the subject matter improved after using the FLoRA platform. & 5-point Likert (1 = No improvement \dots~5 = Significant improvement) & 1.6\% \\ \midrule
\texttt{satisfy\_flora} & Q8: whether students are satisfied with the experience on the FLoRA platform. & 5-point Likert (1 = Very dissatisfied \dots~5 = Very satisfied) & 2.7\% \\ \midrule
\texttt{use\_other\_tools} & Q9: the other tools students choose the finish the assignment. & Free text & 29.0\% \\
\bottomrule
\end{tabularx}
\end{table}

\section*{Technical Validation} 

To ensure the reliability and validity of all data used in this study, we implemented a multi‑step technical validation protocol. Each step addresses a different potential threat to data quality, and together they provide a coherent basis for confidence in both the dataset and the findings derived from it.

\textcolor{black}{First, we assessed the instrumentation validity of the data collection platform. FLoRA~\cite{li2025floraengine} is open-source and has been used in multiple peer-reviewed studies~\cite{fan2025beware,tang2024facilitating,li2023analytics,van2023introduction,osakwe2024measurement,li2024analytics}. Prior validation work has examined FLoRA trace data by triangulating system logs with concurrent think-aloud protocols and eye-tracking evidence~\cite{fan2022towards,lim2021temporal}. In the think-aloud validation, learners' verbal protocols were coded for cognitive and metacognitive processes and compared with time-aligned trace-derived action or process labels. In the eye-tracking validation, gaze patterns over relevant interface areas were examined alongside logged events to check whether trace records corresponded to observable engagement with reading, annotation, writing, and tool-use activities. These studies provide prior support for the instrumentation validity of FLoRA's logging and trace-labelling approach. The present study did not collect new think-aloud or eye-tracking data; therefore, this evidence should be interpreted as prior platform-level validation rather than direct validation of every event in the current dataset.}

\begin{figure}[ht]
\centering
\includegraphics[width=0.9\linewidth]{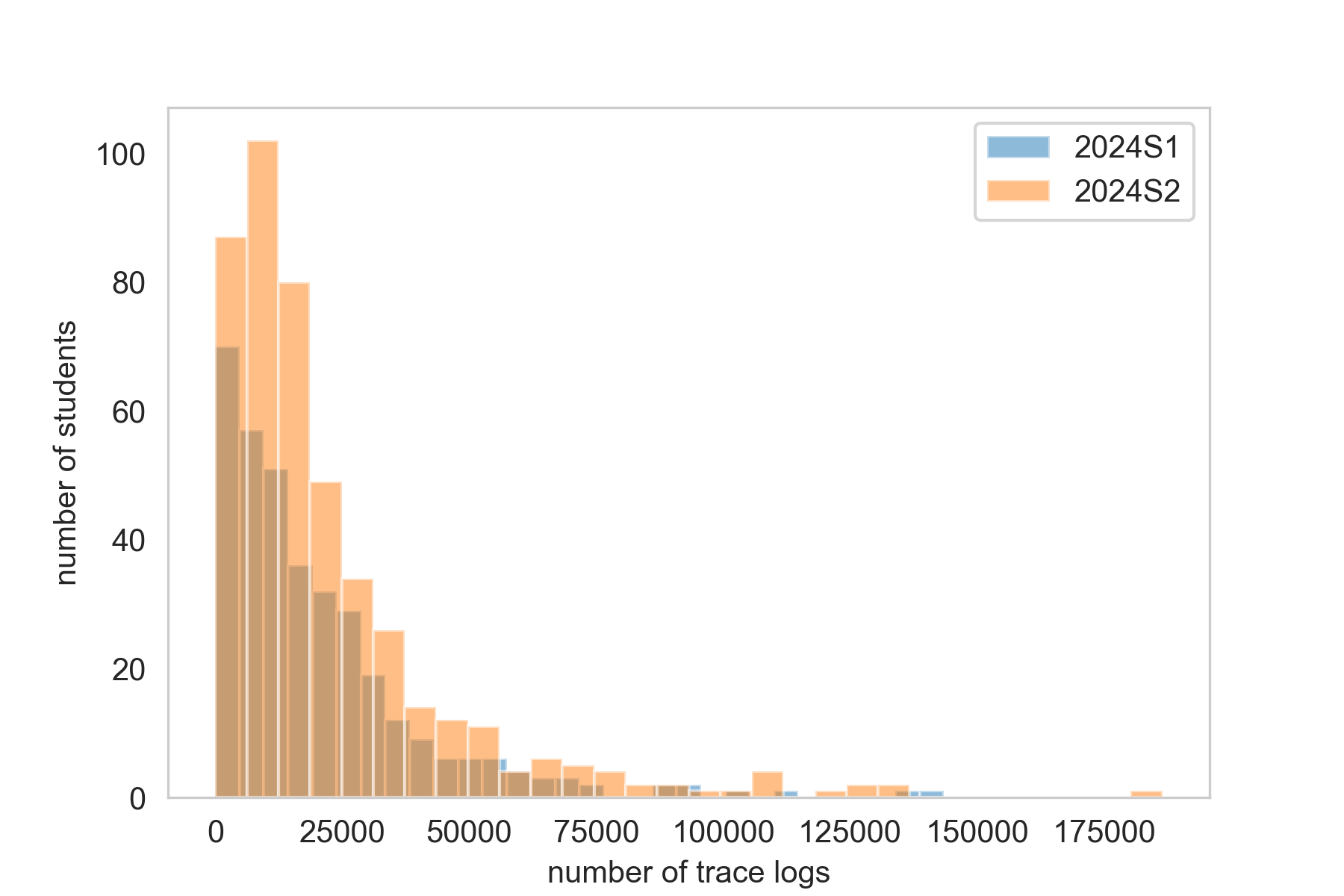}
\caption{Students Interaction Trace Data Distribution}
\label{fig:trace_data_distribution}
\end{figure}

Second, we tested the stability of the trace data across cohorts (temporal stability and replicability). As described in the Methods, the study was conducted in two offerings of the same course: a pilot cohort in 2024S1 and the main cohort in 2024S2 (released as open data). We assessed stability by comparing the distribution of the number of time-stamped raw trace events (log entries) per student between 2024S1 and 2024S2. After conducting normality and homogeneity‑of‑variance tests, we found that both datasets were non‑normal but exhibited equal variances. We therefore used a Mann–Whitney U test, which yielded p = 0.171063 (> 0.05), indicating no significant difference in the distribution of trace logs per student between 2024S1 and 2024S2. Figure~\ref{fig:trace_data_distribution} shows comparable right‑skewed distributions across semesters, with similar medians and a comparable spread of outliers; although 2024S2 includes more students, the distributional patterns remain stable. This consistency supports the reliability of the collection process and the generalisability of our findings.

Third, we quality‑checked all utterance data (content integrity). Trained data raters examined every user question and chatbot response line by line to verify appropriateness (removing any offensive language), relevance to the study context and no non-English content. This review ensured that anomalies or inappropriate items were identified and addressed promptly.

Fourth, we verified the assessment outcome measure (rubric transparency and integrity). As described in the Methods, students’ proposals were graded by experienced course teaching staff using a criterion-based rubric with predefined criteria and weights. Using an explicit rubric provides a structured and transparent learning-outcome measure for linking performance with students’ interactions and trace data in FLoRA.

Finally, we verified survey data integrity. As described in the Methods, the research team reviewed pre-task and post-task survey responses and removed incomplete entries for key variables when appropriate. The complete survey instruments (all questions and response options) are provided in Table~\ref{tab:survey_biographic}, \ref{tab:survey_prior_knowledge}, and \ref{tab:survey_post_task}., enabling replication and reuse.

Together, these procedures triangulate platform validity, cross-semester stability of trace data, dialogue content integrity, rubric-based outcome transparency, and survey data integrity, thereby strengthening the overall credibility of the dataset and the conclusions drawn from it.

\section*{Usage Notes} 

This dataset is best suited for research about IPS, including fine-grained analysis of students’ writing  and revision processes during the synthesis/production stage while interacting with GenAI. Researchers can use the dialogue sequences—together with behavioural and writing logs—to examine indicators of planning, monitoring, evaluation, and adaptation across sessions. Because the interaction log data is pre-labelled with learning-action categories (Table~\ref{tab:learning_action}), it can be used to analyse students' IPS behaviours at a fine temporal granularity. Researchers interested in SRL process-level analysis can further map sequences of learning actions to Bannert-aligned cognitive and metacognitive SRL processes by applying an established process library (see Fan et al.~\cite{fan2022towards}). \textcolor{black}{Once SRL processes are recovered from the trace data using the methodology of Fan et al.\cite{fan2022towards}, they can be linked with chatbot turns and writing actions to operationalise GenAI-specific constructs that have received increasing attention in the literature. For example, sequences in which \texttt{ASK\_CHATBOT} is immediately followed by \texttt{PASTE\_TEXT\_ESSAY} with little intervening monitoring or evaluation process can serve as candidate signatures of \textbf{over-reliance} or \textbf{cognitive offloading}, whereas sequences in which \texttt{READ\_FEEDBACK\_CHATBOT} is followed by evaluation processes and original \texttt{WRITE\_ESSAY} activity reflect productive use of GenAI for \textbf{metacognitive monitoring} and synthesis. Lim et al.\cite{lim2021temporal} and the FLoRA engine papers~\cite{li2025flora,li2025floraengine} provide examples of such temporally-aligned trace analyses that can serve as templates for similar work on this dataset.}

Building on these process measures, the dataset also enables rich GenAI interaction studies and performance modelling. Prompting strategies, query refinement, and response uptake can be characterized using dialogue features such as turn length, intent categories, reformulation frequency, and evaluation moves, which can then be connected to subsequent writing actions in the trace data. These micro‑level behaviours, along with GenAI usage patterns, can be related to macro‑level results such as proposal scores, while controlling for prior knowledge and background survey variables.

The data can also support comparative analyses of different levels or patterns of GenAI engagement and survey-based analyses of students’ perceptions. Because GenAI engagement is observational rather than experimentally assigned, comparisons between low- and high-intensity users, or between different interaction profiles, should be interpreted cautiously and are best examined with appropriate statistical controls or quasi-experimental designs where applicable, such as propensity score matching or inverse probability weighting. In addition, survey responses on perceived GenAI usefulness, engagement, satisfaction, and related constructs can be analysed psychometrically before being related to trace- and dialogue-based behaviours.

\textcolor{black}{The transferability of this dataset should be interpreted at the level of IPS processes and analytic frameworks rather than as direct generalisation of all behavioural patterns to every GenAI system. The dataset was collected in FLoRA, a scaffolded educational platform that integrated an API-based GPT-4o chatbot with annotation and writing tools. Although this configuration differs from open-ended systems such as ChatGPT or Claude, especially when those systems include live web search, retrieval augmentation, citation generation, or external tool-calling functions, the core dialogue-mediated IPS workflow remains comparable: learners formulate information needs, ask questions, read and evaluate responses, refine prompts, seek feedback, and integrate information into written products. Therefore, the dataset can inform future studies of conversational GenAI use by providing a multimodal data schema, dialogue coding framework, IPS-phase mapping, and process-analytic approach linking chatbot interaction with reading, annotation, writing traces, final artefacts, and assessment outcomes.}

\textcolor{black}{At the same time, system-specific capabilities may change how learners search for information, verify sources, rely on generated responses, and translate GenAI support into task performance. For example, web-search-enabled or tool-augmented systems may produce more source-grounded responses, introduce new verification behaviours, or shift students' reliance from internal reasoning to system-provided retrieval outputs. Accordingly, behavioural frequencies, performance associations, and effect-size estimates observed in this FLoRA-based dataset should be replicated before being generalised to other GenAI interfaces, model configurations, disciplines, or educational populations.}

\subsection*{Limitations}

\textcolor{black}{Researchers should account for possible external tool use that is not visible in the FLoRA logs. The dataset records activities conducted within FLoRA and therefore cannot fully capture off-platform drafting, peer support, or interactions with external GenAI chatbots. During preprocessing, cases with missing writing traces or substantial mismatches between FLoRA-composed proposals and Moodle submissions were excluded, which helped remove cases where the proposal appeared to have been mainly written outside FLoRA. However, these steps cannot detect partial drafting in external tools or separate use of GenAI chatbots outside the platform. Such external GenAI use was not technically monitored and was not prohibited by the assignment policy, although students were encouraged to complete the task in FLoRA. The \texttt{PASTE\_TEXT\_ESSAY} action (Table~\ref{tab:learning_action}) and the post-task \texttt{use\_other\_tools} survey item (Table~\ref{tab:survey_post_task}) can be used as proxy indicators or sensitivity variables, but they should not be treated as complete measures of external tool use.}

\textcolor{black}{The dataset should also be interpreted in light of its institutional and disciplinary context. Data were collected at a single public Australian university in a postgraduate Foundations of Data Science course. The sample is therefore most directly generalisable to postgraduate students in computing- and data-science-related programmes, particularly in internationalised higher-education settings. Patterns observed in this dataset may not transfer directly to other disciplines or institutions. For example, students in humanities or social-science courses may approach source evaluation, writing, and GenAI use differently; institutions may differ in students' prior exposure to GenAI and in their policies around AI use; and learners from different linguistic backgrounds may vary in prompting, interpretation, and evaluation strategies. The released survey data include demographic, language-background, and prior-knowledge variables, which can support subgroup and sensitivity analyses, but these variables do not remove the need for replication in other educational contexts.}

\textcolor{black}{A further limitation concerns educational level. The present dataset covers postgraduate learners only, although IPS is relevant across many stages of education. Younger learners may differ from postgraduate students in reading comprehension, information evaluation, academic writing experience, metacognitive monitoring, and familiarity with institutional expectations around AI use. Findings from this dataset should therefore not be assumed to describe undergraduate, secondary-school, or primary-school learners. Future work should adapt the data-collection design to other educational levels using age-appropriate tasks, consent procedures, and scaffolding. The open-source release of FLoRA and the accompanying data schema are intended to support such replications.}

\section*{Data Availability}
The data cleaning code is open sourced at GitHub (https://github.com/Xinyu-Li/FLoRA\_Open\_Data\_IPS) and the data are open sourced at figshare (https://doi.org/10.6084/m9.figshare.32321499).
\section*{Code Availability} 

The FLoRA that is utilised for data collection in this study was developed by our team and is available as open-source software on GitHub (available at https://github.com/Xinyu-Li/Flora). A comprehensive README file is included to provide detailed instructions on deploying and using the platform for data collection. The platform supports a diverse range of data types and experimental scenarios. By adjusting configurations, users can tailor the platform to collect the specific data they require.

\bibliography{sample}


\section*{Funding} 

This research is funded partially by the Australian Government through the Australian Research Council (project number DP240100069 and DP220101209) and the Jacobs Foundation (CELLA 2 CERES).

\section*{Author contributions statement}


X.L., K.Y., Y.C. and G.C. selected the data and conducted data labelling. X.L., K.Y., J.W. and G.C. prepared the dataset for publication. G.C. and D.G. conceptualised and supervised the data collection. X.L. drafted the manuscript. K.Y., J.W., Y.C., G.C. and D.G. reviewed and approved the final version of the manuscript.

\section*{Competing interests} 
The authors declare no competing interests.

\end{document}